# Present status and first results of the final focus beam line at the KEK Accelerator Test Facility

P. Bambade,[1,6,*] M. Alabau Pons,[2] J. Amann,[3] D. Angal-Kalinin,[4] R. Apsimon,[5] S. Araki,[6] A. Aryshev,[6] S. Bai,[7] P. Bellomo,[3] D. Bett,[5] G. Blair,[9] B. Bolzon,[8] S. Boogert,[9] G. Boorman,[9] P. N. Burrows,[5] G. Christian,[5] P. Coe,[5] B. Constance,[5] J.-P. Delahaye,[10] L. Deacon,[9] E. Elsen,[11] A. Faus-Golfe,[2] M. Fukuda,[6] J. Gao,[7] N. Geffroy,[8] E. Gianfelice-Wendt,[12] H. Guler,[13] H. Hayano,[6] A.-Y. Heo,[14] Y. Honda,[6] J. Y. Huang,[15] W. H. Hwang,[15] Y. Iwashita,[16] A. Jeremie,[8] J. Jones,[4] Y. Kamiya,[17] P. Karataev,[9] E.-S. Kim,[14] H.-S. Kim,[14] S. H. Kim,[15] S. Komamiya,[17] K. Kubo,[6] T. Kume,[6] S. Kuroda,[6] B. Lam,[3] A. Lyapin,[18] M. Masuzawa,[6] D. McCormick,[3] S. Molloy,[9] T. Naito,[6] T. Nakamura,[17] J. Nelson,[3] D. Okamoto,[19] T. Okugi,[6] M. Oroku,[17] Y. J. Park,[15] B. Parker,[20] E. Paterson,[3] C. Perry,[5] M. Pivi,[3] T. Raubenheimer,[3] Y. Renier,[1,6] J. Resta-Lopez,[5] C. Rimbault,[1] M. Ross,[12] T. Sanuki,[19] A. Scarfe,[21] D. Schulte,[10] A. Seryi,[3] C. Spencer,[3] T. Suehara,[17] R. Sugahara,[6] C. Swinson,[5] T. Takahashi,[22] T. Tauchi,[6] N. Terunuma,[6] R. Tomas,[10] J. Urakawa,[6] D. Urner,[5] M. Verderi,[13] M.-H. Wang,[3] M. Warden,[5] M. Wendt,[12] G. White,[3] W. Wittmer,[3] A. Wolski,[23] M. Woodley,[3] Y. Yamaguchi,[17] T. Yamanaka,[17] Y. Yan,[3] H. Yoda,[17] K. Yokoya,[6] F. Zhou,[3] and F. Zimmermann[10]

(ATF Collaboration)

[1]*LAL, Université Paris-Sud, CNRS/IN2P3, Orsay, France*
[2]*Instituto de Fisica Corpuscular (CSIC–University of Valencia), Valencia, Spain*
[3]*SLAC National Accelerator Laboratory, Menlo Park, California 94025, USA*
[4]*Cockcroft Institute, STFC, Daresbury Laboratory, United Kingdom*
[5]*John Adams Institute, Oxford, United Kingdom*
[6]*High Energy Accelerator Research Organization, Tsukuba, Japan*
[7]*Institute of High Energy Physics, Beijing China*
[8]*LAPP, Université de Savoie, CNRS/IN2P3, Annecy-le-Vieux, France*
[9]*John Adams Institute, Royal Holloway, United Kingdom*
[10]*European Organization for Nuclear Research, Geneva, Switzerland*
[11]*Deutsches Elektronen-Synchrotron, Hamburg, Germany*
[12]*Fermi National Accelerator Laboratory, Batavia, Illinois 60510-5011, USA*
[13]*Laboratoire Leprince-Ringuet, CNRS/IN2P3, Ecole Polytechnique, Palaiseau, France*
[14]*Kyungpook National University, Korea*
[15]*PAL, Korea*
[16]*Kyoto ICR, Japan*
[17]*The University of Tokyo, Japan*
[18]*UCL, London, United Kingdom*
[19]*Tohoku University, Japan*
[20]*Brookhaven National Laboratory, Upton, New York 11973-5000, USA*
[21]*Cockcroft Institute, University of Manchester, United Kingdom*
[22]*Hiroshima University, Japan*
[23]*Cockcroft Institute, University of Liverpool, United Kingdom*

ATF2 is a final-focus test beam line which aims to focus the low emittance beam from the ATF damping ring to a vertical size of about 37 nm and to demonstrate nanometer level beam stability. Several advanced beam diagnostics and feedback tools are used. In December 2008, construction and installation were completed and beam commissioning started, supported by an international team of Asian, European, and U.S. scientists. The present status and first results are described.

## I. INTRODUCTION

An important technical challenge of future linear collider projects such as ILC [1] or CLIC [2] is the collision of extremely small beams of a few nanometers in vertical size. This challenge involves three distinct issues: creating small emittance beams, preserving the emittance during acceleration and transport, and finally focusing the beams to nanometers before colliding them. The Accelerator Test Facility (ATF) at KEK [3] was built to create small emittance beams, and has succeeded in obtaining emittances that almost satisfy ILC requirements. The ATF2 facility

*Corresponding author.

TABLE I. Comparison of ATF2 parameters with ILC and CLIC specifications.

| Parameters | ATF2 | ILC | CLIC |
|---|---|---|---|
| Beam energy [GeV] | 1.3 | 250 | 1500 |
| L* [m] | 1 | 3.5–4.5 | 3.5 |
| $\gamma\varepsilon_{x/y}$ [m rad] | $5 \times 10^{-6}/3 \times 10^{-8}$ | $1 \times 10^{-5}/4 \times 10^{-8}$ | $6.6 \times 10^{-7}/2 \times 10^{-8}$ |
| IP $\beta_{x/y}$ [mm] | 4/0.1 | 21/0.4 | 6.9/0.07 |
| IP $\eta'$ [rad] | 0.14 | 0.0094 | 0.00144 |
| $\sigma_E$ [%] | ~0.1 | ~0.1 | ~0.3 |
| Chromaticity | $\sim 1 \times 10^4$ | $\sim 1 \times 10^4$ | $\sim 5 \times 10^4$ |
| Number of bunches | 1–3 (goal 1) | ~3000 | 312 |
| Number of bunches | 3–30 (goal 2) | ~3000 | 312 |
| Bunch population | $1-2 \times 10^{10}$ | $2 \times 10^{10}$ | $3.7 \times 10^9$ |
| IP $\sigma_y$ [nm] | 37 | 5.7 | 0.7 |

[4], which uses the beam extracted from the ATF damping ring (DR), was constructed to address the last two issues: focusing the beams to nanometer scale vertical beam sizes and providing nanometer level stability. ATF2 is a follow-up of the final focus test beam (FFTB) experiment at SLAC [5]. The optics of the final focus is a scaled-down version of the ILC design. It is based on a scheme of local chromaticity correction [6] which is now also used for the CLIC design, where symmetries are introduced in the optics to control all relevant aberrations up to third order.

The main parameters of ATF2 are given in Table I with the corresponding values for the ILC and CLIC projects. The value of $\beta_y$ and hence the vertical beam size at the optical focal point [referred to as interaction point (IP) by analogy to the linear collider collision point] are chosen to yield a chromaticity of similar magnitude as in the ILC final focus. For the energy and emittance of the ATF beam and given the distance L* between the last quadrupole and the IP, this leads to a vertical beam size of about 37 nm, including residual effects from higher-order aberrations.

The layout of the ATF/ATF2 facility and the design optical functions of the ATF2 beam line are displayed in Figs. 1 and 2, respectively. The two main project goals are: goal 1—achieving the 37 nm design vertical beam size at the IP by 2010; and goal 2—stabilizing the beam at that point at the nanometer level by 2012.

Achieving the first goal requires developing and implementing a variety of methods to validate the design optics in the presence of imperfections, in particular beam measurement and tuning techniques to cancel unwanted distortions of the beam phase space.

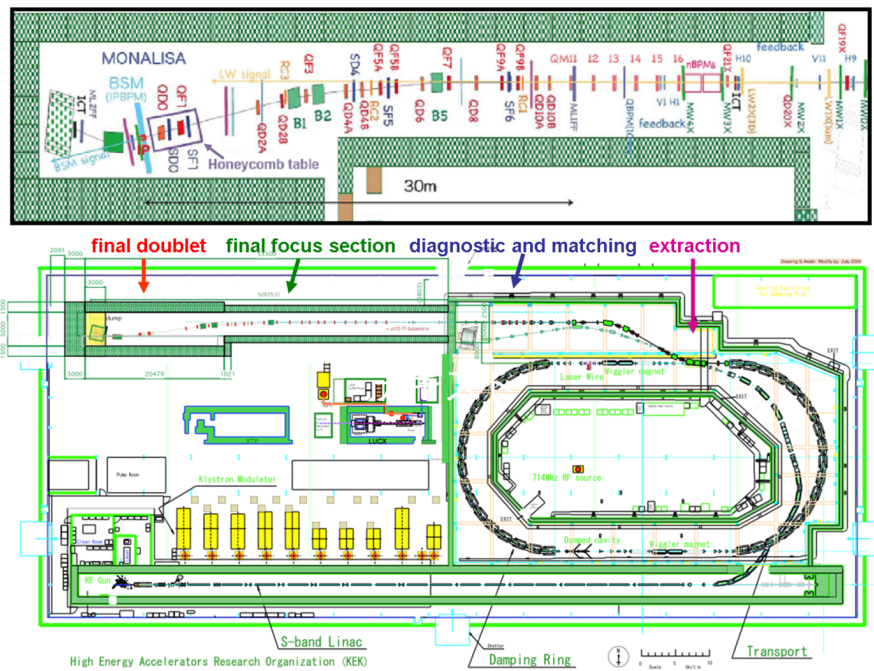

FIG. 1. (Color) Layout of the ATF/ATF2 facility. The second half of the ATF2 beam line is shown in more detail in the upper part of the display.

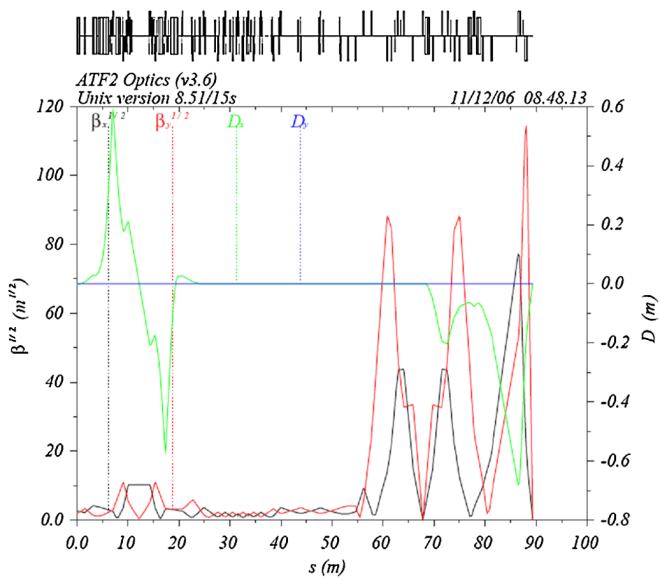

FIG. 2. (Color) ATF2 optics from the ATF damping ring extraction point on the left to the IP on the right. The extraction line extends over the first 25 meters, followed by the diagnostic and matching section in the next 30 meters. The final focus section stretches over the last 35 meters.

Before reaching the ATF2 final focus (see Fig. 1), the beam is extracted from the DR into a reconfigured version of the old ATF extraction line and transported in a matching and diagnostic section where beam parameters can be measured with wire scanners and where anomalous dispersion, betatron mismatch, and coupling can be corrected with a set of dedicated upright and skew quadrupole magnets.

Unlike the case of a linear collider where the measurements of luminosity and electromagnetic interactions between the colliding beams provide information on their respective sizes and overlap, ATF2 is a single beam line. Measuring transverse beam sizes at the IP requires dedicated beam instrumentation, especially a laser interferometer-based beam size monitor (BSM), also called Shintake monitor [7]. Optical adjustments to minimize the beam size at the IP are achieved mainly using combinations of sextupole magnet displacements for independent removal of the linear phase space correlations affecting the beam size.

To measure the beam orbit and maintain the beam size with feedback, the beam line magnets are equipped with submicron resolution cavity beam position monitors (BPM) and are placed on mechanical movers.

Both BSM and BPM measurements are essential to implement the tuning methods for the first goal.

ATF2 construction was completed in 2008 and first beam testing began in December of that year, focusing on the first goal. In addition, a number of studies and hardware development towards the second goal have proceeded in parallel (see Sec. IV). Since the ATF2 project relies on many in kind contributions and is commissioned and operated by scientists from several institutions in a number of countries spread out geographically over three continents, it is considered a model for the organization of the international collaborations which will be needed to build and operate future large scale accelerator projects such as the ILC. Planning and coordination are of crucial importance. The organization of the ATF collaboration and commissioning efforts are described in [3]. The commissioning strategy is designed to use the large international contribution efficiently. Training and transfer of knowledge, important to strengthen the accelerator community and prepare for future large projects, are emphasized. Beam operation time is divided giving 50% for ATF2, 30% for DR and injector related R&D, and 20% for maintenance and upgrades, in order to ensure richness of the overall program while providing sufficient time for the commissioning.

In this paper, the present status and performance of the recently deployed ATF2 systems are described, followed by the first experience with beam measurements and tuning during winter and spring 2009. In the last section, the immediate outlook of the project as well as several near future and longer term plans are outlined.

## II. STATUS OF ATF2 SYSTEMS

### A. Magnets and magnet mover

The ATF2 beam line extends over about 90 meters from the beam extraction point in the ATF DR to the IP (see Fig. 3). It contains seven dipole, three septum, 49 quadrupole, five sextupole, and a number of corrector magnets at room temperature [8,9]. Some magnets were fabricated specially for ATF2 while others were reused from the old ATF extraction beam line and from beam lines at SLAC. Among the latter were the two quadrupole and sextupole magnets composing the final doublet (FD) system at the end of the beam line. The apertures of the FD quadrupole magnets were increased to accommodate the large $\beta$ function values in the FD. Careful magnetic measurements of all newly built magnets and of the modified ones in the FD were done to check and control their higher-order multipole contents. In the last focusing quadrupole magnet within the FD system, where the horizontal beam size reaches its largest value in the system, the tolerances to enable the nominal IP beam parameters to be achieved were slightly exceeded. Several possibilities to readjust the optics design have been studied to mitigate the resulting deterioration of the vertical beam size [10,11].

Dipole and quadrupole magnets in the extraction line were fixed on stainless steel supports bolted to the floor while final focus magnets were fixed on support blocks in concrete glued to the floor with adhesive polymer concrete. Vertical and horizontal positions and tilts of both sets can be adjusted manually with bolts during alignment. Anticipating gradual movements of supports and magnets due to thermal variations or slow ground motion, 20 quadrupole and five sextupole magnets in the final focus were put on remote-controlled three-axis movers recycled from the FFTB experiment. Each mover has three camshafts for adjustments of horizontal and vertical positions (with precision of 1–2 $\mu$m), and for rotations about the beam axis (roll, with precision of 3–5 $\mu$rad). By combining horizontal and vertical motions of these magnets, both trajectory and linear optics distortions can be corrected.

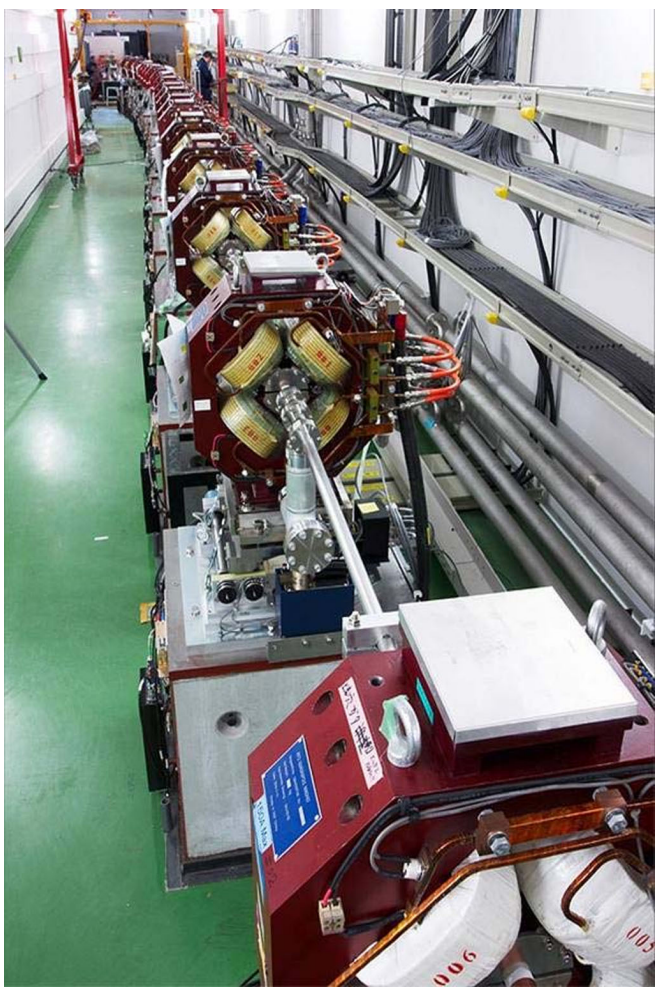

FIG. 3. (Color) View looking downstream along the final focus section of the ATF2 beam line.

Magnet production and refurbishment was carried out from mid-2005 through mid-2008. Installation and alignment was completed in 2008. Overall alignment precisions of 0.1 mm in the three directions and 0.1 mrad in roll angles have been achieved using conventional alignment/metrology techniques. Commissioning and operation of all magnets has proceeded smoothly. The integrated strength data from the magnetic measurements is taken into account in the power supply control software to set required currents and define the standardization procedures. The final alignment of the magnets will be achieved via beam based alignment (BBA) techniques [12].

### B. Final doublet stability

The FD is composed of two quadrupole and two sextupole magnets named QD0, QF1 and SD0, SF1. These magnets must be supported in a way which ensures that their jitter relative to the IP where the BSM is located is smaller than 7 nm, in order to limit effects on the measured beam size to less than 5%. Because of the low beam repetition rate of about 1 Hz, such stability is needed from about 100 Hz, above which ground motion becomes small enough, down to about 0.1 Hz, below which beam based feedback methods can be used. A rigid support was chosen since the coherence length at ATF2, of about 4 m in this frequency range [13], exceeds the distance between the FD and IP, hence strongly suppressing their relative motion. A rigid honeycomb block from Technical Manufacturing Corporation was used, supported on a set of steel plates which covered most of its base and were tied to the floor with bolts. A thin layer of natural beeswax was then used between the plates and the honeycomb block to ensure good mechanical coupling as well as ease of removal [14]. New supports were made and put under the FFTB movers so the magnets' centers reached the 1.2 m beam height. Vibration measurements with the table fixed to the floor and all magnets and movers installed were performed in the laboratory for prior validation, including checking potential effects from cooling water flowing in the magnets [15]. The whole system (see Fig. 4) was installed at KEK in September 2008, where additional measurements [16] were performed confirming that the residual motions of the magnets relative to the BSM were within tolerances (see Table II).

### C. Cavity beam position monitors

The ATF2 beam line is instrumented with 32 C-band (6.5 GHz) and four S-band (2.8 GHz) high resolution cavity beam position monitor systems. In addition to these dipole cavities there are four C-band and one S-band reference cavities to monitor beam charge and beam arrival phase. In the diagnostics and final focus section every quadrupole and sextupole magnet is instrumented with such a BPM. The C-band position sensitive cavities are aligned to the quadrupole magnet centers using mounting fixtures while the S-band cavities are mounted next to the FD magnets. All cavities are cylindrical resonant cavities, with rectangular waveguide couplers to select the dipole, position sensitive, mode. The cavity output is filtered, mixed, amplified, and filtered again to produce an intermediate frequency between 20 and 25 MHz. The down-converted signal is acquired using 100 MHz, 14-bit digitizers with a virtual machine environment interface. The dipole signals are processed using a digital down-conversion algorithm running on a dedicated CPU. The data is acquired, the processing algorithm controlled, and position data distributed via an EPICS channel access application. The radio frequency (rf) electronics and digital down-conversion algorithm are monitored by injection of a suitable frequency triggered rf tone instead of cavity signal. The tone calibration system can be used to monitor the overall electronics and algorithm health without a beam in ATF2.

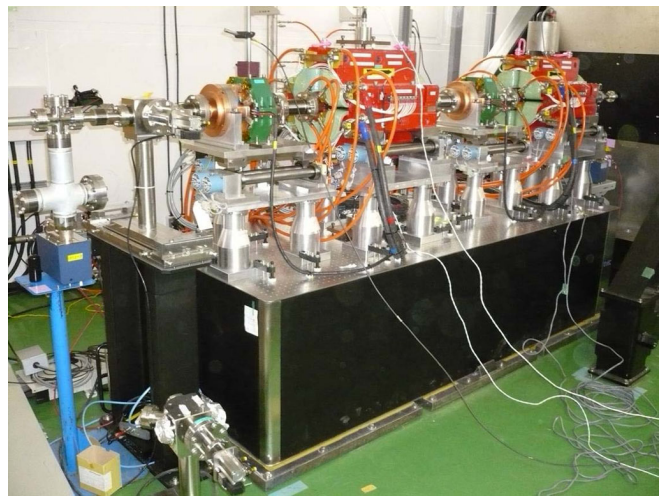

FIG. 4. (Color) View of the final doublet installed on its rigid mechanical support system.

TABLE II. Root-mean square vertical motion of QD0 and QF1 magnets relative to the top of the BSM rigid mount support, for frequencies between 0.2 and 100 Hz. The absolute motion is the coherent movement of both magnets and of the BSM support.

| Magnet | Tolerance (nm) | Relative motion (nm) | Absolute motion (nm) |
|---|---|---|---|
| QD0 | 7 | 5.1 | 212.6 |
| QF1 | 20 | 6.5 | 212.6 |

During the initial ATF2 commissioning, work on cavity BPMs has focused on calibration and on usage of the system for BBA and dispersion measurements. Two types of calibration are required, quadrupole mover calibration and beam deflection using upstream horizontal and vertical corrector magnets (for BPMs on quadrupole magnets without a mover system). The mover calibrations were easy and proved most successful during the initial operation, while the beam deflection calibration suffered from problems with signal saturation and the need to rely on the optics model to propagate beam trajectories. The mover calibration consisted of moving each quadrupole in steps of 100 $\mu$m over a total range of 400 $\mu$m while recording the cavity response for 10 machine pulses at each position. The dynamic range of the cavity system was found to be greater than the range of possible motions of the quadrupole movers ($\pm 1.5$ mm) The resolution of the C-band system is about 1 $\mu$m at this moment, although a full analysis with beam motion jitter subtraction has yet to be performed. More details on the cavity design, fabrication, and performance can be found in [17–19].

### D. Beam size monitor

The beam size monitor used to measure the beam size at the IP is based on inverse Compton scattering between the electron beam and a laser interference fringe pattern [7]. In such a monitor, the energy of the generated gamma rays is typically rather small compared to that of bremsstrahlung photons composing the main background (emitted when beam tail electrons interact with apertures and start showering). In the monitor designed for ATF2 [20], the signal is separated from this high energy background by analyzing the longitudinal shower profile measured with a multilayered detector (located a few meters after the IP after a dipole magnet) [21]. The laser wavelength used is 532 nm, the 2nd harmonic of the Nd:YAG laser, providing a suitable fringe pitch to measure the target vertical size of 37 nm. Four laser beam crossing modes are available to provide a broad dynamic range of up to several microns for the initial beam tuning down to the nominal beam size or less. In addition, a laser wire mode can be used for horizontal beam size measurements.

The system was installed on a rigid mount support [22] at the end of the beam line during summer 2008. After a first checkout with beam in December 2008, commissioning started in 2009 with the laser wire mode. This mode of operation was successfully and reproducibly established during winter and spring runs. The method to properly set up the electron and laser beams was developed experimentally using diagnostics and instrumentation available for both beams. It consists of four main steps: (i) carefully tuning the electron beam trajectory to reduce backgrounds, (ii) aligning the photon detector onto the electron beam axis at the IP, (iii) checking the synchronization of both beams, and (iv) scanning the laser beam horizontally to overlap its waist with that of the electron beam.

Figure 5 shows an example of signal intensity as a function of laser position in the horizontal plane. The relative accuracy of the signal intensity measurement, obtained analyzing the longitudinal profile of the shower in the multilayered photon detector, ranges from 10% to 20%. The line shows a Gaussian fit to the data. The measured horizontal size was 13 microns, consistent with the expectation from folding the 10 microns design waist size of the laser wire with the beam size of about 10 microns available during the run and confirmed by wire scanner measurements downstream of the IP [23].

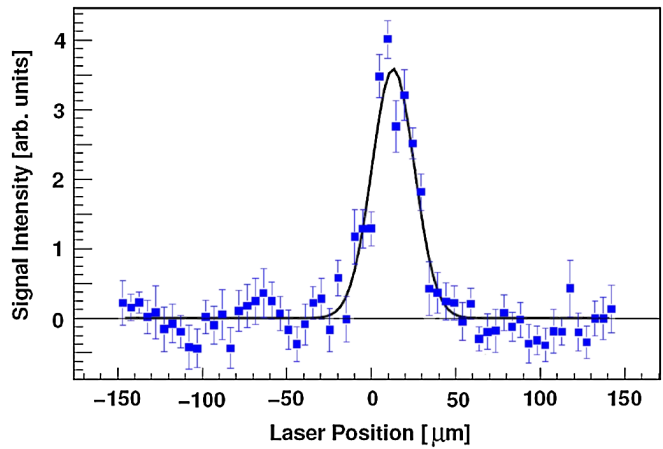

FIG. 5. (Color) Convoluted horizontal size measured by BSM in laser wire mode in March 2009.

### E. Other beam line instrumentation

The instrumentation from the old ATF extraction line [strip line BPMs, ICTs, optical transition radiation (OTR), screen profile monitors, and wire scanners] is reused in the reconfigured beam line. There are five wire scanners with tungsten and carbon wires of 10 and 7 $\mu$m diameter, respectively, located in the diagnostic section upstream of the final focus section (see Fig. 1). They are used to measure the horizontal and vertical beam emittances after extraction from the DR. An additional wire scanner is installed just downstream of the IP for beam size tuning and has tungsten and carbon wires of, respectively, 10 and 5 $\mu$m in diameter. Screen monitors are located right after the extraction, in the middle of the beam line, and before and after the FD. An optical fiber beam loss monitor is installed all along the beam line to localize and quantify beam losses in a relative sense.

### F. Beam line modeling tools

The successful tuning of the ATF2 beam line relies on many automated software tools prepared and tested throughout the collaboration. To facilitate broad participation in the corresponding tasks, a "Flight Simulator" software environment was designed as a middle layer between the existing lower level ATF control system based on EPICS and V-system and the higher-level beam dynamics modeling tools [24]. This is a "portable" control system for ATF2 that allows code development and checkout offsite and additionally provides the framework for integrating that code into the operational ATF2 control system. The software developed through the flight simulator is developed mainly through the LUCRETIA [25] package while various "add-on" packages are also supported to enable usage of MAD8, PLACET, and SAD [26–28] optics programs. It is used in the ATF2 control room alongside tools developed through the existing V-system interface. Tools currently in use include: extraction line coupling correction, extraction line dispersion measurement and correction, extraction line and final focus orbit monitoring and steering, optical tuning knobs for the IP spot size based on moving sextupole magnets, BPM display and diagnostics tools (orbit plotting, reference save/restore system, offline calibration of strip-line BPMs), watchdog tools (e.g. monitoring of the beam orbit in critical apertures, magnet strengths, online optics checks, model response matrix), magnet standardization, orbit bump and BBA tools to extract the offsets between BPMs and quadrupole or sextupole magnet centers.

## III. FIRST ATF2 BEAM MEASUREMENT AND TUNING RESULTS

### A. Overview of commissioning runs

Since the beginning of commissioning at the end of December 2008, five commissioning runs were dedicated to ATF2, each two or three weeks long. In the December 2008 run, only some of the magnets were turned on, in a configuration with a large $\beta^*$. The beam was brought to its dump with minimal beam losses to pass a radiation inspection required at KEK and to enable basic hardware and software checks. The following runs in February–March 2009 also used a large $\beta^*$ (8 cm horizontally and vertically), this time with all ATF2 magnets switched on for the first time and an optical configuration with basic features similar to the nominal optics [29]. For such values of the $\beta^*$ parameters, respectively 20 and 800 times larger than the nominal values, beam sizes in the FD are reduced by the square root of the same factors. This was important to ease requirements for backgrounds while producing IP beam spots with $\sigma_{x,y} \sim 12.5, 1-2$ $\mu$m which were measurable by the BSM in laser wire mode or were just below the resolution limit of the tungsten post-IP wire scanner. In the most recent April and May 2009 runs, the vertical $\beta^*$ was reduced to 1 cm, corresponding to an IP spot of $\sigma_y \sim 0.5$ $\mu$m. As the chromaticity is not yet predominant for such a value, sextupole magnets had little influence and could be turned off. In parallel with the gradual deployment of software control tools and continuous testing and characterization of the BSM and of the cavity and strip line BPMs, first measurements of the optical functions and beam parameters were also pursued.

### B. Beam tuning strategy

Focusing the low emittance beam extracted from the ATF DR to the specified IP beam size requires correcting trajectory and optics distortions induced both by imperfections along the beam line and by mismatch of the beam phase space at DR extraction. While final corrections must be done at the IP, it is still important to keep mismatches under control at the entrance of the final focus, in order to limit distortions of the linear optics in the carefully tuned chromatic correction section and to minimize backgrounds in the BSM from bremsstrahlung, which can be emitted and reach the detector when beam tail particles reach the vacuum chamber at high-$\beta$ points of the optics and start showering.

The beam tuning sequence followed in successive shifts during April and May 2009 runs was (i) bring the beam to the dump with maximal transmission using the chosen magnet configuration and flatten the trajectory, (ii) successive BBA in selected "critical" quadrupole magnets, (iii) dispersion measurement in the diagnostic and matching section, followed by correction using the upright and skew quadrupole magnets in the extraction line, (iv) emittance and Twiss parameter measurements combined with coupling correction with the system of dedicated skew quadrupole magnets in the diagnostic and matching section, (v) horizontal and vertical waist-scan and dispersion measurements with the FD set to focus the beam at the post-IP wire scanner, in order to infer $\beta^*$, (vi) if needed, rematch $\beta^*$ to its target values using dedicated quadrupole magnets immediately upstream of the final focus section, (vii) vertical beam spot minimization at the post-IP wire scanner by canceling residual dispersion and coupling, using orthogonal combinations of vertical motions of sextupole magnets in the final focus and, alternatively, the set of upstream skew quadrupole magnets, (viii) reset the FD for IP focusing to enable BSM measurements, and (ix) if backgrounds are too large in the BSM, rematch $\beta^*$ in the horizontal plane to larger values.

### C. Extracted vertical beam emittance and betatron matching

Vertical emittances of less than 10 pm were consistently achieved in the DR during spring 2009 [30]. After extraction to ATF2, several effects can however enlarge it, especially anomalous dispersion and coupling remaining from the DR or generated in the extraction process. In the March 2009 run and during earlier tests in 2007–2008 [31] before reconfiguring the extraction line for ATF2, large growth factors were often observed. In April and May, systematic BBA in selected quadrupole magnets of the extraction line, followed by careful corrections for residual dispersion and coupling, enabled the reproducible measurement of vertical emittance values in the 10 to 30 pm range. Figure 6 shows results from one of the measurements during May 2009. The horizontal and vertical emittances were 1.7 nm and 11 pm, respectively, with rather good horizontal matching but some apparent mismatch vertically, presumably partly due to not having fully corrected the residual coupling.

### D. Measurements of first-order optics at IP

In the large $\beta^*$ optics used, IP beam sizes are essentially determined by the first-order optical transfer matrices, higher order effects being negligible. Beam size measurements at the IP can thus serve to check the first-order optics, by comparing with nominal values or propagating Twiss parameters measured upstream. Waist scans can be done after setting the FD to focus the beam in both planes at either the BSM or post-IP wire scanner, using orthogonal

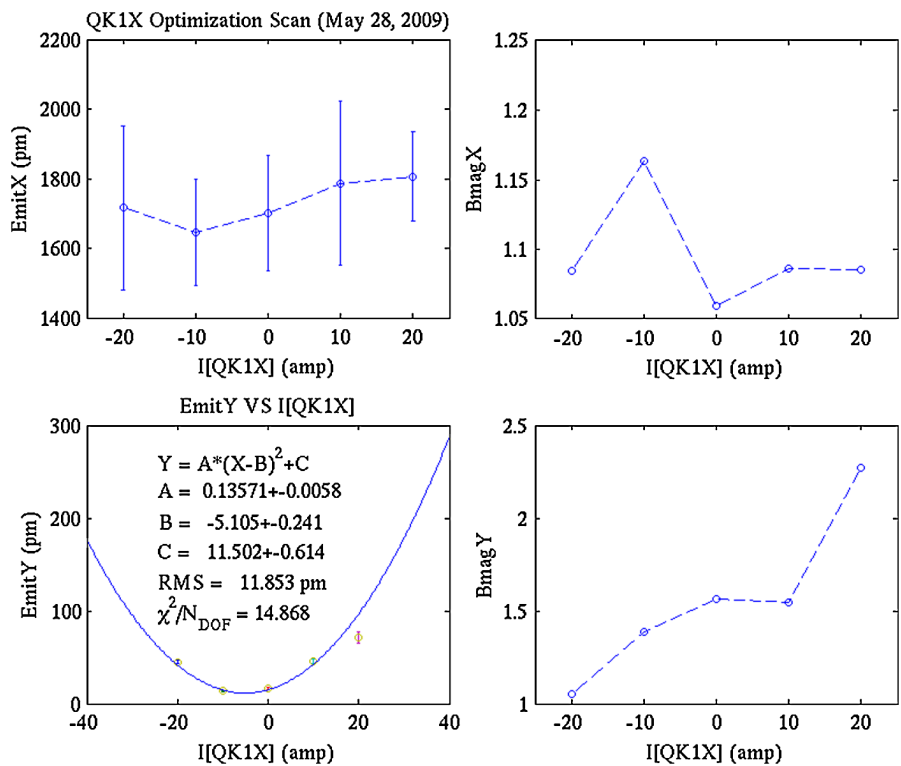

FIG. 6. (Color) Emittances and betatron mismatch factors (Bmag X, Y) measured in the ATF2 extraction line in May 2009 during coupling minimization with one of the skew correction quadrupoles. The Bmag X, Y factors quantify the distortions of the beam ellipses in both transverse planes with respect to those of the ideal matched betatron phase space, following the definition in [43]. In the vertical dimension, BmagY exceeds unity indicating apparent mismatch.

QD0, QF1 combinations (for independent control in each plane) or just QD0. From the parabolic dependence of the square of the beam size with respect to quadrupole magnet strength, the emittance and Twiss parameters can be computed. The values obtained are however biased if the beam size at the minimum of the parabola is below the instrumental resolution or if there is significant residual dispersion or coupling.

In the horizontal plane, the nominal beam size of about 12.5 $\mu$m could be measured with both the post-IP tungsten wire scanners and BSM laser wire mode. Since the horizontal emittance is much larger than the vertical one, residual coupling has a negligible effect. However, there is significant horizontal dispersion in the nominal optics near the IP (see Fig. 2), resulting from the local chromaticity correction scheme, which must be accounted for along with any residual mismatch propagated from imperfectly corrected upstream errors. Figure 7 shows an example of horizontal waist scans from May 2009 [32]. The extracted emittance and $\beta^*$ values at the minimum, $\varepsilon_x = (1.13 \pm 0.06)$ nm and $\beta_x = (13 \pm 1)$ cm, could be compared with the values expected at the post-IP wire scanner in the design optics (2 nm, 10 cm) as well as with the ones obtained propagating the measurements made in the extraction line in a previous shift (1.7 nm, 14.5 cm). A full analysis to evaluate the significance and possible origin of differences has yet to be performed.

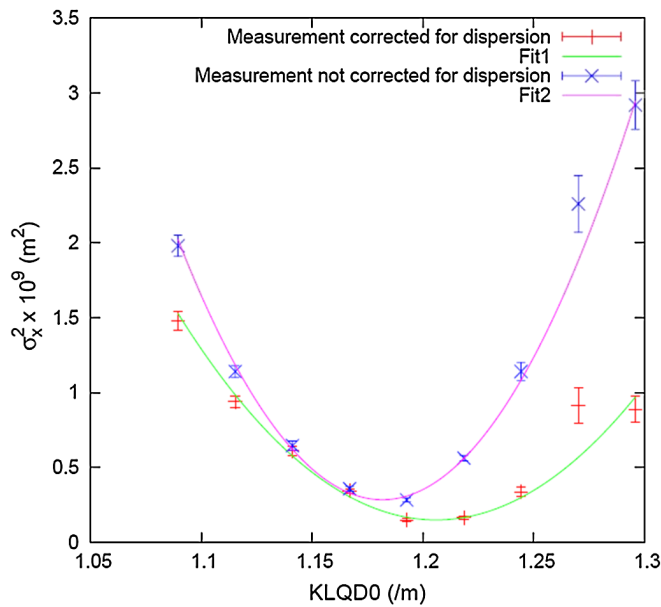

FIG. 7. (Color) Horizontal beam size squared measured at the post-IP wire scanner in May 2009 as a function of QD0 strength. Blue crosses are before and red crosses are after correcting for the measured horizontal dispersion. Green and pink lines are the corresponding parabolic fits. The shift in abscissa between corrected and uncorrected parabolas is due to anomalous horizontal dispersion.

Similar waist scans were recorded in the vertical plane using the 10 $\mu$m diameter post-IP tungsten wire scanner. Since the expected minimum value could not be resolved in these scans, they were used only to evaluate the divergence of the beam near the IP. Combining with the emittance measured in the matching section, estimates of $\beta_y$ could be obtained in the assumption of negligible residual coupling. Data recorded in the last week of May 2009 gave, for instance, $\beta_y \sim 10$ mm after correcting for anomalous vertical dispersion measured during the scan, somewhat less than the nominal value expected at the post-IP wire scanner (18 mm).

### E. Vertical spot minimization at the IP

To cancel residual vertical dispersion and coupling at the IP, the best results obtained so far were with a procedure consisting of sequentially using six skew quadrupole corrector magnets (as well as a pair of normal quadrupole magnets in the extraction and matching sections) to empirically reduce the vertical beam spot size at the post-IP wire scanner down to about 3 $\mu$m, corresponding to the resolution limit of the 10 $\mu$m diameter tungsten wire which was used.

## IV. ATF2 OUTLOOK AND PLANS

The present ATF2 efforts of the ATF collaboration are focused on the first ATF2 goal. The priority for the next runs during autumn 2009 is to measure submicron vertical beam sizes using the interference mode of the BSM. This will involve continued operation with the large $\beta^*$ optics (8 cm horizontally and 1 cm vertically) in order to confirm its properties in more detail. Vertical dispersion and coupling effects at the IP will need to be corrected well enough to reduce the vertical beam size to below about 1 $\mu$m, corresponding to the resolution limit of the 5 $\mu$m carbon wire scanner behind the IP. The tuning sequence outlined in the previous section will be followed.

During the summer shutdown in 2009, a number of improvements were made, especially in the BSM and BPM systems. The magnets in all the beam lines of the ATF facility were also realigned.

The BSM will use a laser crossing angle of 4.5 degrees, for which the sensitivity is maximal for vertical beam sizes of about 1 $\mu$m. A new 3 times more powerful laser will enhance the signal significance with respect to the background. Additional collimation in front of the photon detector will also help to reduce the background from bremsstrahlung emitted upstream. Moreover, wire scanner, screen, and knife edge monitors were newly installed in the IP chamber of the BSM to make it easier to overlap the electron and laser beams.

The performance of the cavity BPMs was extensively studied to characterize and improve stability and reproducibility in signal amplitudes and phases over long periods of up to a month. The electronics of the strip line BPMs is also being upgraded to suppress residual kicker noise picked up on the electrodes, shown to degrade performances, and to enable more reliable calibration.

During 2010, the goal will be to reduce the $\beta^*$ parameters enough towards the nominal values for vertical beam sizes smaller than 100 nm to be measured. Preparations towards this goal are on-going in parallel with the above tasks. In particular, a new system with multiple OTR stations is being prepared in the diagnostic section to supplement the existing wire scanners and enable speedier and more precise 2D profile measurements. The improved Twiss parameter, emittance, and coupling determinations which will result should help to minimize the extracted vertical emittance. Two additional tasks related to the BSM are also important to achieve the first goal in 2010: improved automation of the BSM data acquisition along with integration into the overall software environment for beam size tuning at the IP, and evaluation and control of BSM beam induced backgrounds, in particular as a function of $\beta^*$.

The ATF collaboration also pursues several other hardware developments of particular relevance to future linear colliders, especially in the context of the second ATF2 goal: characterization of the site and beam line stability [33], the MONALISA interferometer system [34] for accurate monitoring of the FD position with respect to that of the BSM, the feedback on nanosecond time scale (FONT) project [35], the nanometer resolution IP-BPM project [36], the fast nanosecond rise time kicker project [37], and a new cavity-BPM optimized to monitor angular variations of the beam near the IP with high accuracy [38]. A laser wire system operated in the old ATF extraction line during 2005–2008 with the aim to demonstrate 1 $\mu$m resolution beam size measurements [39] has also been moved to a new location in the ATF2 diagnostics section for further testing and development in coming years. In the future, this system could be expanded to replace some or all present wire scanners. Future linear colliders are expected to rely extensively on laser wire systems, so it is important to gain experience operating a multiple system in realistic conditions.

Plans to upgrade the performance of ATF2 on the time scale of a few years, after the main goals of ATF2 have been achieved, are also under consideration. In particular, optical configurations with ultralow $\beta^*$ values (2 to 4 times smaller than nominal in the horizontal and vertical planes), relevant to both the CLIC design and to some of the alternative ILC beam parameter sets [1], are actively studied [2]. There is also a proposal to upgrade the FD with superconducting magnets [40] built according to ILC direct wind technology, to allow stability studies with beam of direct relevance to the setup planned at ILC. An R&D program to develop a tunable permanent magnet suitable for the FD is also pursued in parallel, with as an initial goal the construction of a prototype for initial beam testing in the upstream part of the ATF2 beam line [41]. Since possibilities to achieve the smallest vertical beam sizes are limited, especially in the case of reduced $\beta^*$ values, both by the field quality in the magnets of the presently installed FD [10,11] and by their aperture (to avoid excessive bremsstrahlung photon background in the BSM), these proposals are naturally connected in the sense that an upgraded FD should also aim to both enlarge the aperture and improve the field quality.

Longer term, more tentative, plans being discussed include, after 2012, the possibility of a photon facility, with laser and optical cavities for the planned photon linear collider and generation of a photon beam. Strong QED experiments with laser intensities of $>10^{22}$ W/cm$^2$ could then also be considered, e.g., to pursue experimental studies of the predicted Unruh radiation [42].

## V. CONCLUSION

The ATF collaboration has completed the construction of ATF2 and has started its commissioning. Important experience operating the new cavity BPM and BSM instrumentation in real conditions has been gained and first beam measurements have been performed in a magnetic configuration with reduced optical demagnification. Both horizontal and vertical emittances were successfully tuned and measured in the extraction line, with values approaching the design values of 2 nm and 12 pm, respectively. First checks of the first-order optics along the beam line and at the IP were also done. Hardware developments for the second ATF2 goal are being pursued in parallel with the present commissioning work for the first goal. The collaboration is also preparing several near and long-term plans for ATF2. In the next few years, information very valuable for any future collider with local chromaticity correction and tuning of very low emittance beams can be expected. In the previous experience at the FFTB, the smallest vertical beam sizes which were achieved were about 70 nanometers. The work described here continues to address this largely unexplored regime in a systematic way.

## ACKNOWLEDGMENTS

This work is supported by DOE Contract No. DE-AC02-76SF00515; Grant-in-Aid for Creative Scientific Research of JSPS (KAKENHI 17GS0210); U.S.A.–Japan Collaboration Research Grant of MEXT; Agence Nationale de la Recherche of the French Ministry of Research (Programme Blanc, Project No. ATF2-IN2P3-KEK, Contract No. ANR-06-BLAN-0027); The "Toshiko Yuasa" France Japan Particle Physics Laboratory; Science and Technology Facilities Council, U.K., and EuCARD project cofunded by the European Commission within the Framework Program 7, under Grant Agreement No. 227579.